# High Photoresponsivity and Short Photo Response Times in Few-Layered WSe$_2$ Transistors

Nihar R. Pradhan[1],* Jonathan Ludwig[1,2], Zhengguang Lu[1,2], Daniel Rhodes[1,2], Michael M. Bishop[1], Komalavalli Thirunavukkuarasu [1], Stephen A. McGill[1], Dmitry Smirnov[1] and Luis Balicas[1,2],*

[1]National High Magnetic Field Lab, Florida State University, 1800 E. Paul Dirac Dr. Tallahassee, FL 32310.

[2] Department of Physics, Florida State University, Tallahassee, Florida 32306, USA

**ABSTRACT**. Here, we report the photoconducting response of field-effect transistors based on three atomic layers of chemical vapor transport grown WSe$_2$ crystals mechanically exfoliated onto SiO$_2$. We find that tri-layered WSe$_2$ field-effect transistors, built with the simplest possible architecture, can display high hole mobilities ranging from 350 cm$^2$/Vs at room temperature (saturating at a value of ~500 cm$^2$/Vs below 50 K) displaying a strong photocurrent response which leads to exceptionally high photo responsivities up to 7 A/W under white light illumination of the entire channel for power densities $p < 10^2$ W/m$^2$. Under a fixed wavelength of $\lambda$ = 532 nm and a laser spot size smaller than the conducting channel area we extract photo responsivities approaching 100 mA/W with concomitantly high external quantum efficiencies up to ~ 40% at room temperature. These values surpass values recently reported from more complex architectures, such as graphene and transition metal dichalcogenides based heterostructures. Also, tri-layered $WSe_2$ photo-transistors display photo response times in the order of 10 microseconds. Our results indicate that the addition of a few atomic layers considerably decreases the photo response times, probably by minimizing the interaction with the substrates, while maintaining a very high photo-responsivity.

**INTRODUCTION**. Although graphene[1] is the most studied two-dimensional, or layered material, single or few atomic layers of other van der Waals solids, such as insulating $h$-$BN$[2] or semiconducting transition metal dichalcogenides (TMDs) such as $MoS_2$,[3,4] $MoSe_2$,[5,6,7] or $WSe_2$,[8,9] are being intensively investigated, as either promising gate dielectrics or as the conducting channel material in field-effect transistors (FETs). Since monolayer $MoS_2$ has been shown to be a direct-bandgap semiconductor,[10,11] due to quantum-mechanical confinement[10,12,13], it was suggested that it could be a suitable material for optoelectronic applications[14] where the direct bandgap would naturally lead to a high absorption coefficient and therefore to efficient electron–hole pair generation under photo-excitation. Its potential for applicability is further supported by the overall behavior of the FETs, which display high ON to OFF current ratios, i.e. up to $10^8$,[4] photo-gains as high as 5000,[15] and high specific detectivities, i. e. up to $10^{10}$-$10^{11}$ Jones.[16]

However, the photoresponse of single-layer $MoS_2$ is a matter of controversy, with reports indicating either the absence of a sizeable response[16,17] or, and in contrast, a high photocurrent when the phototransistor is its "ON" state.[14,18,19,20] Most reports attribute the observed photocurrent to the separation of photogenerated electron and hole pairs at the interface between the semiconductor and the metallic electrical contact[14-20] although others claim a prominent role for the photothermoelectric effect (resulting from the difference in the Seebeck coefficients between the contact and the channel materials and the presence of light-induced temperature gradients).[21] The Schottky barriers between the semiconducting channel and the metallic electrodes, which result from the difference between the electron affinity of the semiconductor and the work function of the metal, are expected to play a major role in any conventional photodetector.[22] Several groups tentatively attributing the differences among the observed electrical and/or optical transport properties, to the size of the Schottky barrier.[15,18,21] In addition, the bending of the conduction and valence bands in the vicinity of the metallic contacts, which leads to pronounced local electric fields under a bias voltage, were claimed to contribute significantly to photocurrent generation.[20]

The electrical contacts would also play a relevant role for the photo-conducting temporal response. It is reported to vary widely from characteristic rising/decay times inferior to a 1 ms to values surpassing 30 s, for either mechanically exfoliated or chemical vapor deposited (CVD) $MoS_2$.[15-18] For instance, phototransistors based on large-area, CVD synthesized, $WSe_2$ monolayers electrically contacted with Pd, which presumably display small Schottky-barriers, are found to exhibit high photo gain ($10^5$) and specific detectivity ($10^{14}$ Jones), but with characteristic response times > 5 s in air.[23] In contrast, when these phototransistors are contacted with Ti, which is claimed to lead to a higher Schottky-barrier, they would display a fast response time, i.e. shorter than 23 ms, but with several orders of magnitude lower photo gain and specific detectivity.[23]

Here, we explore the photoresponse properties of field-effect transistors based on three atomic-layers of $WSe_2$ exfoliated onto $SiO_2$:$p$-Si substrates contacted with a single combination of Ti:Au contacts. We find large photoresponsivities approaching 7 A/W (under white light and for illumination power densities $10 < p < 10^2$ W/m$^2$) and external quantum efficiencies approaching 40 % (under laser illumination at λ = 532 nm). These values can still increase considerably by varying either the bias or the back gate voltages or both, and/or simply by illuminating the whole area of the conducting channel. Their characteristic response time is found to remain within a few tenths of microseconds. Improvements in the quality of the electrical contacts, the starting materials and substrates is likely to improve their performance making few-layer WSe$_2$ phototransistors potentially useful for optoelectronics applications.

**RESULTS AND DISCUSSION**

Figure 1**a** shows a schematic of our multi-layered WSe$_2$ based field-effect transistors and the configuration of measurements.-The devices are composed of three atomic layers of WSe$_2$, which are electrically contacted by using a combination of Ti and Au. Figure 1**b** shows a micrograph of one of our tri-layered WSe$_2$ FETs whose detailed electrical transport characterization is presented in the Supporting Information and which will be briefly discussed below. As seen in figure 1**c,** the spot size of the laser beam (3.5 μm) was intentionally chosen to be smaller than the area of the conducting channel in order to minimize the interaction with the area surrounding the electrical contacts. Figures 1**d** displays an atomic force microscopy (AFM) image of the edge of the WSe$_2$ from which we extracted the height profile shown in figure 1**e**.

Our tri-layered WSe$_2$ FETs behave as hole-doped transistors, i.e. yielding a sizeable current (in the order of 1 μA/μm at low excitation voltages $V_{ds}$) only for negative gate voltages $V_{bg}$ surpassing a certain threshold value $V^t_{bg}$, see Figure 2. The ON to OFF ratio surpasses 10$^6$ but with relatively large sub-threshold swings SS ≅ 5 V. The room temperature values of $V^t_{bg}$ is sample dependent varying from -30 and -40 V, and increasing as the temperature is lowered, as can be seen by comparing figures 2**a** and **c** and **b** and **d**. This behavior is attributable to disorder-induced carrier localization in the channel as extensively studied in Si/SiO$_2$ MOSFETs.[24] Above $V^t_{bg}$ and from the slopes of $I_{ds}$ as function of $V_{bg}$, and by applying the MOSFET transconductance formula, $\mu_{Fe} = 1/c_g \, d\sigma/dV_{bg}$, where $\sigma = I_{ds} \ell / V_{ds} w$ with $\ell$ and $w$ being the length and the width of the channel respectively, and $c_g = \varepsilon_r \varepsilon_0 / d = 12.789 \times 10^{-9}$ F/cm$^2$ for a $d = 270$ nm thick SiO$_2$ layer, we obtain maximum two-terminal field-effect hole mobilities exceeding $\mu_{FE} \cong 350$ cm$^2$/Vs at $T = 275$ K. $\mu_{FE}$ increases to values surpassing ~500 cm$^2$/Vs at $T = 20$ K due to the suppression of phonon scattering, see also supplemental figure S1. These near room-temperature values for the *two-terminal* field-effect mobilities are considerably higher than the ones extracted for electron-doped multi-layered MoS$_2$ FETs.[25] This suggests that our chemical vapor transport synthesized WSe$_2$ single-crystals are less disordered than natural MoS$_2$. Our extracted mobilities are higher than those obtained at $T = 80$ K for encapsulated multi-layered WSe$_2$ using graphene for the electrical contacts and ionic liquid gating[26] but close to the value of ~600 cm$^2$/Vs extracted also for ionic liquid gated and *h*-BN encapsulated multi-layered WSe$_2$ at $T = 220$ K.[27] Insofar the highest room temperature mobility reported for multilayered WSe$_2$, i.e. ~ 500 cm$^2$/Vs, was obtained by using parylene as the gate dielectric.[28] On SiO$_2$ we obtained $\mu_{FE}$ ~ 350 cm$^2$/Vs at $T = 300$ K for crystals composed of approximately 10 atomic layers, and whose value increased up to ~ 670 cm$^2$/Vs at $T = 105$ K.[29] Most importantly, in these samples we found a broad agreement between field-effect and Hall mobilities, with their difference ascribed to an underestimation of the value of the gate capacitance used in the transconductance formula, most likely due to the presence of spurious charges.[29] The gate-induced decrease in the size of the Schottky barriers at the contacts and the screening of both impurities and charge carriers in the neighborhood of the mobility edge should contribute to higher field-effect mobilities.

As discussed above, the metallic electrical contacts were claimed to play a major role in the photoconducting response of FETs based on TMDs, therefore we proceed to analyze the quality of our Au (~90 nm) on Ti (~4 nm) contacts and evaluate the size of the Schottky barriers. As seen in figure 3**a** at low excitation voltages $V_{ds}$, the drain to source current $I_{ds}$ is linear (or ohmic like) in $V_{ds}$, independently of the applied back gate-voltage $V_{bg}$. This indicates that any Schottky barrier between WSe$_2$ and the contacts plays a quite mild role at room temperature, with carriers being promoted across the barrier by thermionic emission or thermionic field emission processes.[30,31] In effect, according to thermionic emission theory,[30,31] the drain to source current $I_{ds}$ is related to the Schottky barrier height $\phi_{SB}$ through the expression:

$$I \cong AA^*T^2 \exp\left(\frac{e\phi_{SB}}{k_B T}\right)\left[1 - \exp\left(-\frac{eV_{ds}}{k_B T}\right)\right] \qquad (1)$$

where, $A$ is the area of the Schottky junction, $A^* = 4\pi e m^* k_B^2 h^{-3}$ is the effective Richardson constant, $e$ is the elementary charge, $k_B$ is the Boltzmann constant, $m^*$ is the effective mass and $h$ is the Planck constant. In order to evaluate the Schottky barrier at the level of the contacts, in figure 3**c** we plot $I_{ds}$ normalized by the square of the temperature $T^2$ as a function of $e/k_B T$ and for several values of $V_{bg}$. Linear fits, depicted by straight lines from which we extract $\phi_{SB}$, are limited to higher temperatures since at lower $T$s one observes pronounced gate dependent deviations from linearity. Figure 3**d** shows $\phi_{SB}$ as a function of $V_{bg}$ where the red line is a linear fit. From the deviation from linearity (the so-called flat band condition) one extracts the size of the Schottky barrier $\Phi$.[31] A pronounced deviation from linearity is observed beyond $V_{bg} \cong -80$ V, where the size of $\Phi$ is of just a few meV indicating a better band alignment than what is a priori expected from the difference between the work function of Ti (4.33 eV) and the electron affinity of WSe$_2$ ($\cong$ 4.00 eV).[32] Therefore, Ti would seem to lead to a small Schottky barrier in contrast to the claims of Ref.[23] The use of a different temperature pre-factor, i.e. $T^\alpha$ with $\alpha = 1.5$ or 1 but $\neq 2$ yields similar results. This small $\Phi$ value might result from defects and impurities[33] around the contact area which would pin the Fermi level at an arbitrary position relative to the conduction and valence bands. Or it might result from thermally assisted tunneling through a quite thin Schottky barrier.

Having clarified this aspect, we now return to the issue of carrier localization due to the disorder in the channel. In Figure S2 we plot the two-terminal conductivity σ as function of $T^{-1/3}$, the dependence expected for the two-dimensional variable range hopping conductivity mechanism. We find that it perfectly describes the data on a broad range of temperatures and gate voltages, but particularly at lower gate voltages, as previously seen by other groups working on FETs based on TMDs.[34,35] We ascribe the threshold back gate voltage $V^t_{bg}$ to the existence of a mobility edge. Atomic level roughness, dangling bonds, and buried charges, for example from Na treatment of the SiO$_2$ layer, create a random Coulomb potential on the SiO$_2$ substrate, contributing to disorder which leads to carrier localization. Dislocations in the TMD layers,[34,35] adsorbates and residues from the fabrication process also contribute to disorder and hence to localization. The initial accumulation of charges in the channel through the field-effect contributes to the screening of spurious charges which act as traps for charge carriers. Additional carriers accumulated into the channel become mobile and yield sizeable currents only when these traps are fully screened, or when $V_{bg} > V^t_{bg}$. $V^t_{bg}$ is temperature-dependent because at higher temperatures thermal activated processes would contribute to carrier detrapping and hence would decrease the values of $V^t_{bg}$, as seen experimentally. At the moment we do not have a detailed understanding on the nature and on the role of disorder in these systems, but our experimental results are unambiguous.

Next, we proceed to evaluate photo-transport properties of our tri-layered WSe$_2$ FETs under coherent light illumination. The transition towards a direct band-gap is predicted to occur, and has indeed been observed in single atomic layers of TMDs.[14] Hence, in monolayers illumination leads to the efficient creation of electron-hole pairs because it does not require the intervention of phonons. Nevertheless, we chose to work with tri-layered devices for two reasons: i) similarly to the motivation for fabricating *h*-BN encapsulated samples[26,27], one would expect the top atomic-layer to act as a capping layer thus minimizing the role of adsorbates and ii) the bottom layer, which is in direct contact with the SiO$_2$ substrate, might be too affected by disorder, spurious charges and concomitant localization to a point where most of the current would be effectively carried by the middle layer.[36] Here, our goal is to evaluate if such architecture would lead to an increase in the photo-generated electrical currents since the usefulness of the TMDs for high performance applications such as photodetectors, will likely depend on whether their current carrying capabilities can rival those of current silicon based devices/technologies.

Figure 4**a** displays the drain to source current $I_{ds}$ as a function of the bias voltage $V_{ds}$ under several values of illumination power from a λ = 532 nm laser (spot size 3.5 um). This data was collected in absence of gate voltage ($V_{bg}$ = 0 V). As seen, $I_{ds}$ displays a non-linear dependence on $V_{ds}$ similarly to what was reported by Ref.[14] on MoS$_2$. The room temperature generated photocurrent, i.e. $I_{ph}$ = $I_{ds}(P)$ - $I_{ds}(P = 0)$, is the difference between the currents measured in dark condition (red trace) and under illumination. We did *not* detect significant time dependence for the photo-generated current once the laser light was turned on and kept at a constant power level. The resulting photo responsivity $R$, which is the ratio between $I_{ph}$ and the applied illumination power $P$, is displayed in Fig. 4b as a function of the excitation voltage $V_{ds}$, and for several values of $P$. As seen, the induced photocurrent increases considerably with increasing $P$. $R$ tends to display an asymmetric dependence on the sign of the excitation voltage $V_{ds}$ (see, supplemental Fig. S 4). As discussed in the Supplemental Information file, the origin of this asymmetry can be ascribed to the relative and non-intentional proximity of the laser spot to one of the current contacts. In effect, as discussed in detail in Ref.[20] the profile of the valence and conduction bands in TMDs changes considerably in the vicinity of the contact area due to the accumulation of charges induced by the Schottky barriers, particularly under the application of a bias voltage $V_{ds}$. The bias voltage further increases the local band bending and consequently it facilitates, under illumination, the promotion of charge carriers through the metallic contacts. In contrast, when $V_{ds}$ is reversed, the concomitant band bending becomes unfavorable for, for example, hole extraction or electron injection into the channel, thus decreasing the extracted photocurrent as experimentally seen. In Ref.[20] it is shown that most of the $I_{ph}$ is generated precisely at the interface between the channel and the electrical contacts where the electrostatic potential is found to be the strongest.

Nevertheless, the most important observation in Fig. 4b is the size of the extracted photo responsivity $R = I_{ph}/P$ *in absence of any gate voltage*. $R$ approaches 150 mA/W, at low $P$s due to photo generated currents ranging from $10^{-9}$ to $10^{-8}$ A, these values are similar to the ones reported in Refs.[16,21,23,37] for several TMDs. It is important to emphasize that the area of our conducting channel is nearly ~27 times larger than the laser spot size, implying that the illumination of the entire channel (under the same power density) should increase the generated photocurrent by one order magnitude or even more. In Figure 4c we plot the concomitant external quantum efficiency EQE, or the number of photo-generated carriers circulating through any given photodetector per adsorbed photon and per unit time, as a function of $V_{ds}$. Here, EQE = $hcI_{ph}/e\lambda P$, where $P$ is the illumination power irradiated onto the channel, $\lambda$ is the excitation wavelength, $h$ is Planck's constant, $c$ is the velocity of light, and $e$ is the electronic charge. Notice how, in absence of gate voltages and under low illumination power, one can extract EQE values up to 40 % under $P$ = 10 nW. However, as will be discussed below, the application of a small gate voltage, in the order of just $V_{bg}$ = - 10 V, increases the value of $I_{ph}$ considerably indicating that it is possible to improve these values considerably by just varying the gate voltage SiOor simply by increasing the bias voltage. Notice that these $R$ values are similar to those reported for multi-layered WS$_2$,[37] although in Ref.[37] the illumination area is 58 μm$^2$, or approximately 6 times larger than our spot size, and $R$ was obtained under a considerably smaller power density. As will be subsequently demonstrated, by increasing the illumination area, under white light in this case, at a fixed power density, one can extract far more pronounced $R$ values, which in combination with appropriate values for $V_{bg}$ and $V_{ds}$ is likely to generate photocurrents approaching (or surpassing) 10$^{-6}$ A. Again we remind the reader that we chose to limit the spot size of our laser beam to minimize the interaction with the contacts in an effort to evaluate the intrinsic photo-responsivity of WSe$_2$.

Figures 4 **d**, **e** and **f** show respectively, $I_{ph}$, $R$ and the resulting EQE as a function of the laser illumination power *P*, under zero gate voltage, for two values of the bias voltage, $V_{ds}$ = +1 and +2 V, and for two crystals of similar thicknesses, i.e. sample #1 (black markers) and #2 (blue markers). In Figure 4**d** red line is a linear fit which yields $I_{ph} \propto P^{0.5}$. The non-linear dependence of $I_{ph}$ on $P$ discards the photothermoelectric effect as the origin of the observed photoresponse. Notice how $R$ and EQE reach maximum values approaching ~ 0.15 A/W and 40 % respectively,

under a modest power of ~ 10 nW, and for $V_{ds}$ = 1 V and $V_{bg}$ = 0 V. This $R$ value is ~ 3 times larger than those extracted from CVD grown multi-layered $WS_2$,[38] and from graphene-$WSe_2$ heterostructures.[39] It is higher or comparable to those extracted for multi-layered $MoS_2$[16,18] despite the very small channel area illuminated by the laser beam, or the fact that the dependence of $R$ (or of EQE) on the excitation wavelength $\lambda$ has yet to be evaluated. As seen in Figure 4**e**, and similarly to what was reported by other groups,[14] the photoresponsivity decreases by more than one order of magnitude as $P$ increases. Most likely, this reflects the shortening of exciton recombination times as $P$ increases due to the concomitant increase in the density of photo-generated electron-hole pairs. Consequently, the external quantum efficiency as seen in Figure 4**f**, is observed to decrease as $P^{-0.5}$, reflecting the $P$-dependence of $I_{ph}$.

Now we proceed to evaluate the photoresponsivity of our tri-layered $WSe_2$ phototransistor under the white light spectrum generated by a Xe lamp when it *illuminates the entire channel*. Figure 5**a** displays the photocurrent $I_{ph}$ as function of the bias voltage, for several back gate voltages under a constant low illumination power density $p$ = 75 W/m². Figure 5**b** on the other hand shows $I_{ph}$ as a function of $V_{ds}$ under $p \cong 4.3$ kW/m². Notice how an increase in $p$, by nearly two orders of magnitude, leads to an increase in $I_{ph}$ by just a little more than one order of magnitude. Figure 5**c** illustrates the evolution of $I_{ph}$ as function of $V_{ds}$ on $p$ under a gate voltage $V_{bg}$ = -12 V. Figure 5**d** on the other hand plots $I_{ph}$ as a function of the gate voltage for several values $p$. As seen, $I_{ph}$ increases considerably as the value of the negative gate voltage increases. This reflects the progressive displacement of $V_{bg}$ towards the threshold gate voltage where the charges become fully delocalized (see Fig. 2). Not surprisingly, this indicates that charge localization (see Fig. S2 and related discussion), due for example, to interface roughness is detrimental to the photoresponsivity of $WSe_2$. Figure 5**e** displays the photoresponsivity as a function of $p$ for several values of the gate voltage. Remarkably, at higher gate voltages and under the lowest values of $p$, which are far below the sunlight power density, one extracts $R$ values surpassing 7 A/W. More importantly, this value can be increased by increasing both the bias and the gate voltages. Such high $R$ values, when compared to the previous results collected under the $\lambda$=532 nm laser beam discussed in Figure 4, can be attributed to i) a larger illumination area and ii) to illumination of the area surrounding the contacts which is likely to contribute to the promotion of charge carriers across the Schottky barriers. Therefore, and although multi-layered transition-metal dichalcogenides are characterized by an indirect band gap requiring the intervention of a phonon to generate an exciton, they are still characterized by pronounced photo responsivities. Although, for our tri-layered crystals excitons associated with the direct band gap are quite likely to play a major role in their photoresponse. In fact, this is further supported by Fig. S3 which shows the photo responsivity of a 12-layer $WSe_2$ crystal indicating $R$ values surpassing 5 A/W at low power densities when collected under a bias voltage of just 0.1 V. In this sample, by increasing $V_{ds}$ to ~ 1 V, as applied to the tri-layered sample, one could increase $I_{ph}$ and the concomitant $R$ values, by at least one order of magnitude.

In supplemental figure S**5**, we show that the photoresponse of our tri-layered $WSe_2$ FETs is frequency-dependent: shortening the excitation wavelength to $\lambda$ = 405 nm does not lead to an overall increase in their photoresponsivity, as one would expect from the observations in Ref.[14]. In fact, it leads to a sharp decrease in $R$ and in the concomitant EQE due perhaps to excitation-dependent characteristic relaxation pathways with a predominant role for non-radiative decay.[40]

One important technological aspect for any potential photodetector is its characteristic temporal response. For example, single layer $MoS_2$ phototransistors displaying high photoresponsivities particularly at very low excitation powers,[14] displayed rather large characteristic response times, i.e. within several seconds.[14,17] But other groups reported a total photocurrent generation/annihilation time of just 50 ms.[18] To characterize the photoresponse dynamics of our tri-layered WSe$_2$ FETs, we used two techniques : (i) on/off light modulation to measure rise and fall time constants, (ii) AC sinusoidal modulation of the excitation light intensity to measure the characteristic response as a function of frequency. Figure 6 **a** shows a schematic of the experimental set-up for measuring time dependent photoresponse. The laser diode output ($\lambda$ = 532 nm) was modulated through a function generator and controlled with a fast Si photodetector. The time-dependent generated photocurrent was converted into a time-dependent voltage, through an *I/V* converter, and captured either by a digitizer, or by a Lock-In amplifier. The measurements were performed at $V_{bg}$ = - 10 V, implying that the FET is in its OFF state without light illumination. Figure 6 **b** plots the photocurrent $I_{ph}$ measured as a function of time with turning the laser on and off. The time constant $\tau_{on/off}$ defined as the time required to change the signal amplitude between 10% and 90%, are about 40 µs for both rising and fall edges. Figure 6 c shows the frequency dependence of the normalized photocurrent measured under sinusoidal light modulation, $I_{ph}$ (*f*) / $I_{ph}$ (*f* →0). The photocurrent begins to decrease beyond a characteristic roll over frequency $f_0$ = 30.7 kHz corresponding to a characteristic transient time $\tau_0 = (2\pi\, f_0)^{-1}$ = 5.2 µs. If instead one followed the bandwidth convention by choosing the frequency at which the photocurrent decreases by 3 dB one would obtain $f_{3dB} \cong$ 43.3 kHz corresponding to a characteristic time constant $\tau_{3dB} = (2\pi f_{3dB})^{-1} \cong$ 3.7 µs. The current amplifier used in our setup is characterized by a rise time of 10 µs and a 3dB bandwidth of 60 kHz corresponding to a time constant of 2.7 µs. Thus, the reported values $\tau_{on/off} \cong$ 40 µs and $\tau_{3dB} \cong$ 3.7 µs represent a conservative estimate, implying that the actual photoresponse of these tri-layered WSe$_2$ FETs is expected to be even faster. These values are comparable to the transient time constant of ~ 5 µs extracted for $SnS_2$ thin crystal arrays which is claimed to be the fastest measured so far for transition metal dichalcogenides based photo-transistors.[41] Also, it is important to emphasize that such short time constants discard any role for the photo-thermoelectric effect. The study in Ref.[23] suggests that the metallic contacts and concomitant Schottky barriers play a determinant role in the photoresponsive times of transition metal dichalcogenide phototransistors. Nevertheless, when using Ti for the electrical contacts, in photodetectors based on monolayer $WSe_2$, they obtain photoresponsive raise times of ~23 ms, or orders of magnitude higher than the photoresponsive times extracted here. This suggests that the interaction with the substrates is detrimental to the photoresponse times of transition metal dichalcogenides monolayers, hence justifying our choice of multilayered crystals.

**CONCLUSIONS**

In summary, we have shown that few layer *p*-doped $WSe_2$ field-effect transistors built with the simplest possible architecture, can display high carrier mobilities and a strong photocurrent response. This leads to a large photo-responsivity (e.g. ~ 7 A/W) and to remarkable external quantum efficiencies (e.g. 40%), surpassing those of Si photodetectors.[42] Most importantly, they also display fast photoresponse times ranging from a few µs to a few tenths of µs depending on the precise definition of the response time. Although much higher photo-responsivities were claimed for single layered $MoS_2$,[14] the values reported here surpass photo-responsivities and external quantum efficiencies extracted from more complex architectures which use graphene for the electrical contacts in heterostructures containing multi-layered transition metal dichalcogenides.[39,43,44] In tri-layered $WSe_2$, photo-responsivities $R > 7$ A/W under white light illumination and transient times ranging from ~4 to ~40 µs represent a good compromise between Ref.[14] which reports, $R > 10^2$ A/W but with associated photo response times of a few seconds, and Ref.[41] which report transient times of 5 µs but with $R < 10$ mA/W.

We emphasize that there is still ample room for improvement of WSe$_2$ based phototransistors. Although our results do indicate that one is already able to extract sizeable photocurrents, a systematic study on the thickness, spectral dependence, role of the electrical contacts and substrates, including bias- and gate-voltage dependences should lead to the extraction of even higher photocurrents than the values reported here, requiring the development of robust contacts. The important point is that the architecture implemented here is rather simple and does not require, for example, the use of dielectric engineering or the fabrication of van der Waals heterostructures,[39,43,44] which increase the complexity and therefore the production costs of potentially commercial phototransistors/photodetectors. Finally, our results are promising enough to justify a major engineering effort focused on increasing the carrier mobility, decreasing the channel and contact capacitances to shorten their characteristic photo-responsive times, and on increasing the carrier tunneling probability across the electrical contacts. If successful, this effort could unleash a new era in ultra-compact, low power, and perhaps even flexible optoelectronics.

## AUTHOR INFORMATION

Corresponding Author
*Email: balicas@magnet.fsu.edu
 Email: pradhan@magnet.fsu.edu

## METHODS

WSe$_2$ single crystals were synthesized through a chemical vapor transport technique using either iodine or excess Se as the transport agent. Multi-layered flakes of WSe$_2$ were exfoliated from these single crystals by using the micromechanical cleavage technique, and transferred onto *p*- doped Si wafers (with doping levels ranging from 2 x $10^{19}$ cm$^{-3}$ to 2 x $10^{20}$ cm$^{-3}$) covered with a 270 nm thick layer of SiO$_2$. For making the electrical contacts 90 nm of Au was deposited onto a 4 nm layer of Ti *via* e-beam evaporation. Contacts were patterned using standard e-beam lithography techniques. After gold deposition, we proceeded with PMMA lift off in acetone. The devices were annealed at 300 $^0$C for ~ 3 h in forming gas, followed by high vacuum annealing for 24 hours at 120$^o$C. Atomic force microscopy (AFM) imaging was performed using the Asylum Research MFP-3D AFM. Electrical characterization was performed by using a combination of sourcemeter (Keithley 2612A) coupled to a Physical Property Measurement System. The photo-response was measured with a homemade micro-optical setup based on a modified microscope (Olympus BX51) enabling photo-current, photoluminescence or Raman measurements. For photo-current measurements under monochromatic illumination, we used a 532 nm solid-state CW laser (Coherent Sapphire 532-150 CDRH) or a 405 nm/ 532 nm laser diode system (Thorlabs DJ532-40,532 nm; DL5146-101S, 405 nm; LTC100-A controller). The laser beam was injected into a single-mode optical fiber, delivered to a microscope sample stage through a 20X achromatic objective and focused into a spot of about 3.5 um in diameter. The incident optical power was adjusted with neutral density filters or by controlling the operating current of the laser diodes. The photo-current was detected with and *I/V* converter (DL1211 current amplifier). To measure the photo-response times, the laser diodes were modulated using an external function generator while the photo-response was measured with a high speed digitizer (NI PXI-5122) or a high-frequency lock-in amplifier (SRS SR844). A Xe lamp was used when performing measurements under white light. The incident illumination power was controlled by using neutral UV-VIS filters placed between the lamp and the sample. An aperture was used to define a spot diameter of ~9.3 mm. A broadband OPHIR-3A detector was used to measure the illumination power density.

*Conflict of Interest:* The authors declare no competing financial interest.

*Acknowledgement.* This work was supported by the U.S. Army Research Office MURI Grant No. W911NF-11-1-0362. J. L. acknowledges the support by NHMFL UCGP No. 5087. Z.L .and D.S. acknowledge the support by DOE BES Division under grant no. DE-FG02- 07ER46451. The NHMFL is supported by NSF through NSF-DMR-0084173 and the State of Florida.

*Supporting Information available:* Drain to source current as a function of back gate voltage for several temperatures, conductivity as function to the inverse of the temperature to the 1/3 power, indicating two-dimensional variable-range hopping conductivity, photocurrent data under white light for a 12 layer device, asymmetry of the photoresponse which is attributable to the interaction between carriers and light around the contact area, and photoconductivity, photo responsivity and external quantum efficiencies for a shorter monochromatic wavelength. This material is available free of charge *via* the Internet at http://pubs.acs.org.

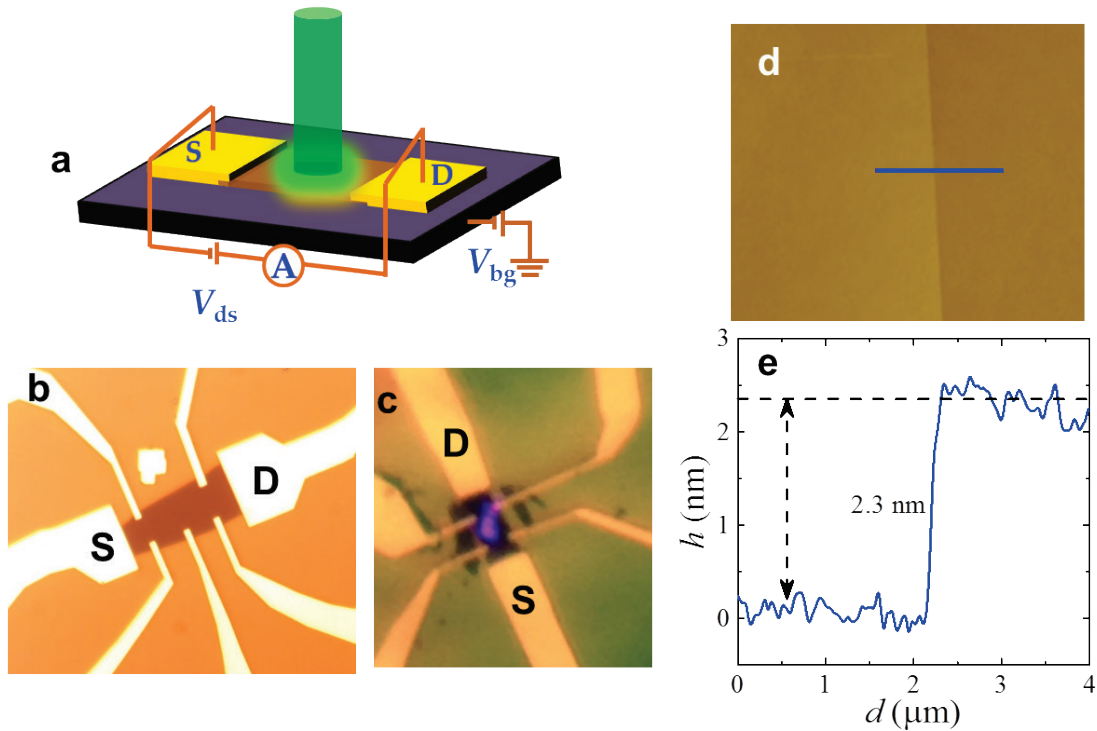

**Figure 1**. (**a**) Sketch of our experimental set-up showing drain (D) and source (S) current electrodes. Green cylinder depicts the laser beam, whose spot size is chosen to be considerably smaller than the active WSe$_2$ channel area to minimize the interaction with the charge carriers flowing through the current leads. (**b**) Micrograph of one of our tri-layered WSe$_2$ based field-effect transistors, indicating electrical current pads, as well as electrical connections for four-terminal resistivity and Hall-effect measurements. This sample of channel length $\ell$ = 22.22 µm and an average width $w$ = 8.37 µm was mainly used for two- terminal electrical transport measurements. Here "S" and "D" denote source and drain contacts, respectively. (**c**) Micrograph of a second tri-layered WSe$_2$ field-effect transistor. This sample of channel length $\ell$ = 15.8 µm and an average width $w$ = 8.6 µm was mainly used for evaluating the photoconducting response. As seen, a violet laser spot is shone onto the channel. (**d**) Atomic force microscopy image along the edge of the sample in (**c**). Blue line indicates the line along which the height profile (shown in **e**) was collected. The height profile indicates a step of 2.3 nm or 3 atomic layers (each with a thickness of 0.647 nm)

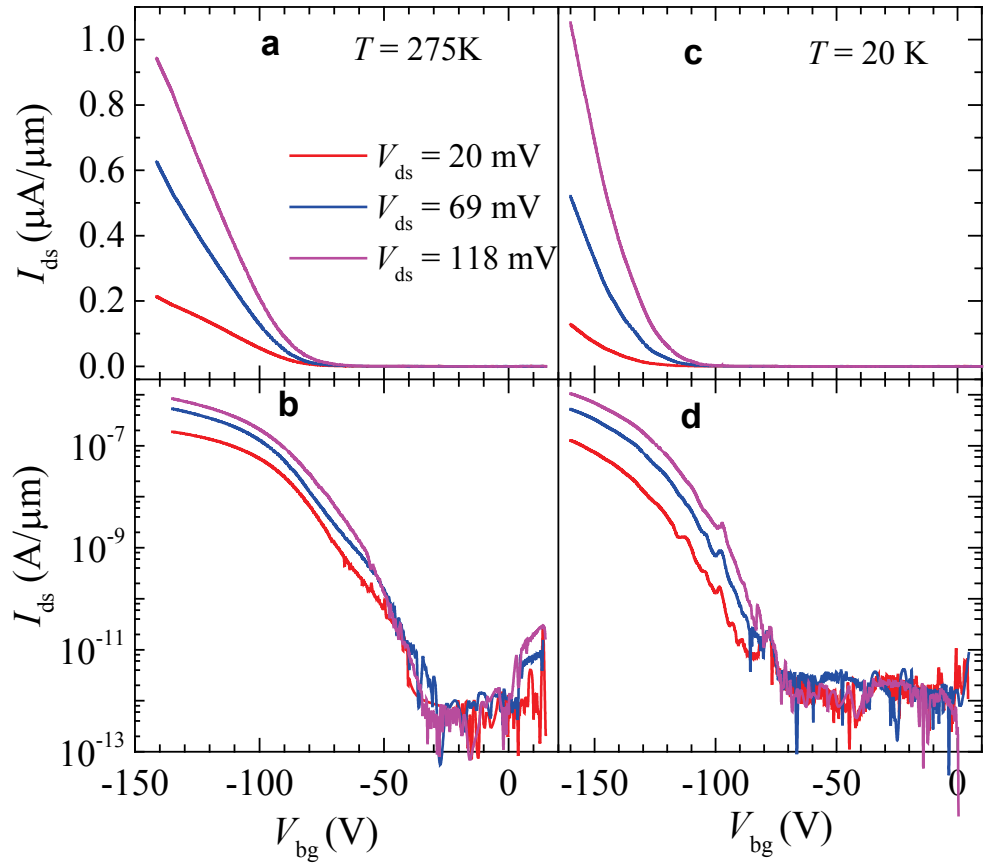

**Figure 2**. (**a**) Source to drain current $I_{ds}$ normalized by the width $w$ of the channel as a function of the back gate voltage $V_{bg}$ and for several values of drain-source excitation voltage $V_{ds}$ of 20 (red line), 69 (blue line ) and 118 mV (magenta line), respectively. As seen a sizeable current $I_{ds}$ is observed only for negative values of $V_{bg}$, indicating hole-conduction once the $V_{bg}$ surpasses a threshold value. (**b**) Same as in **a** but in a logarithmic scale. (**c**) Same as in **a** (including excitation voltages and respective line colors) but at a temperature $T = 20$ K. (**d**) Same as in (**b**) but in a logarithmic scale.

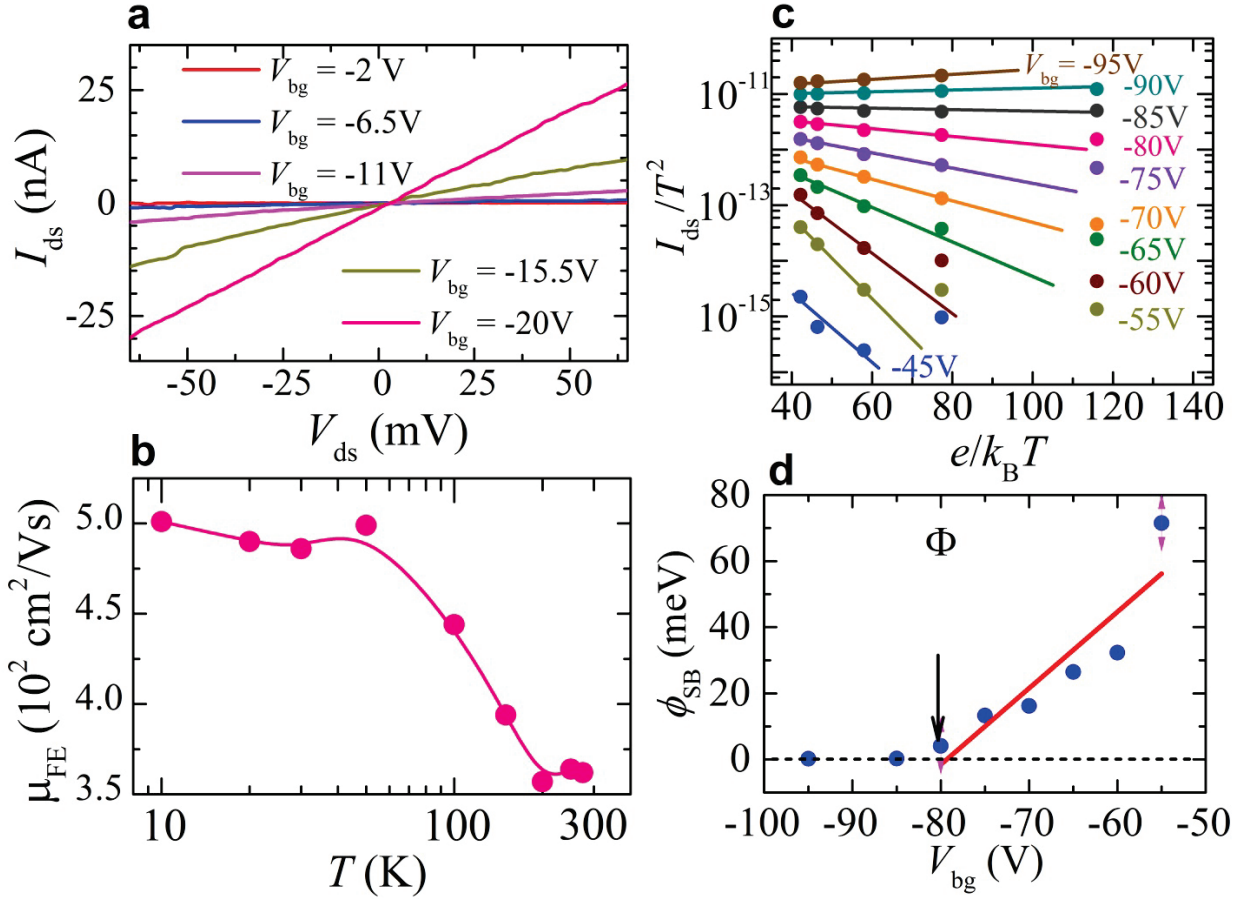

**Figure 3. (a)**. Drain to source current $I_{ds}$ as a function of the excitation voltage $V_{ds}$ for several values of the gate voltage $V_{bg}$. Linear dependence at low excitation voltages a small role for the Schottky barriers. **(b)** Field effect mobility $\mu_{FE}$ as a function of the temperature $T$, as calculated from the curves in Figure S2. For this tri-layered sample, $\mu_{FE}$ saturates at ~ 500 cm$^2$/Vs at low $T$s. **(c)** $I_{ds}$ normalized by the square of the temperature $T^2$ as a function of $e/k_BT$, where $e$ is the electron charge and $k_B$ is the Boltzmann constant. **(d)** $\phi_{SB}$ in as a function of $V_{bg}$. Red line is a linear fit. Deviations from linearity are observed above $V_{bg} \cong -80$ V suggesting a rather small Shottky barrier of just a few meV.

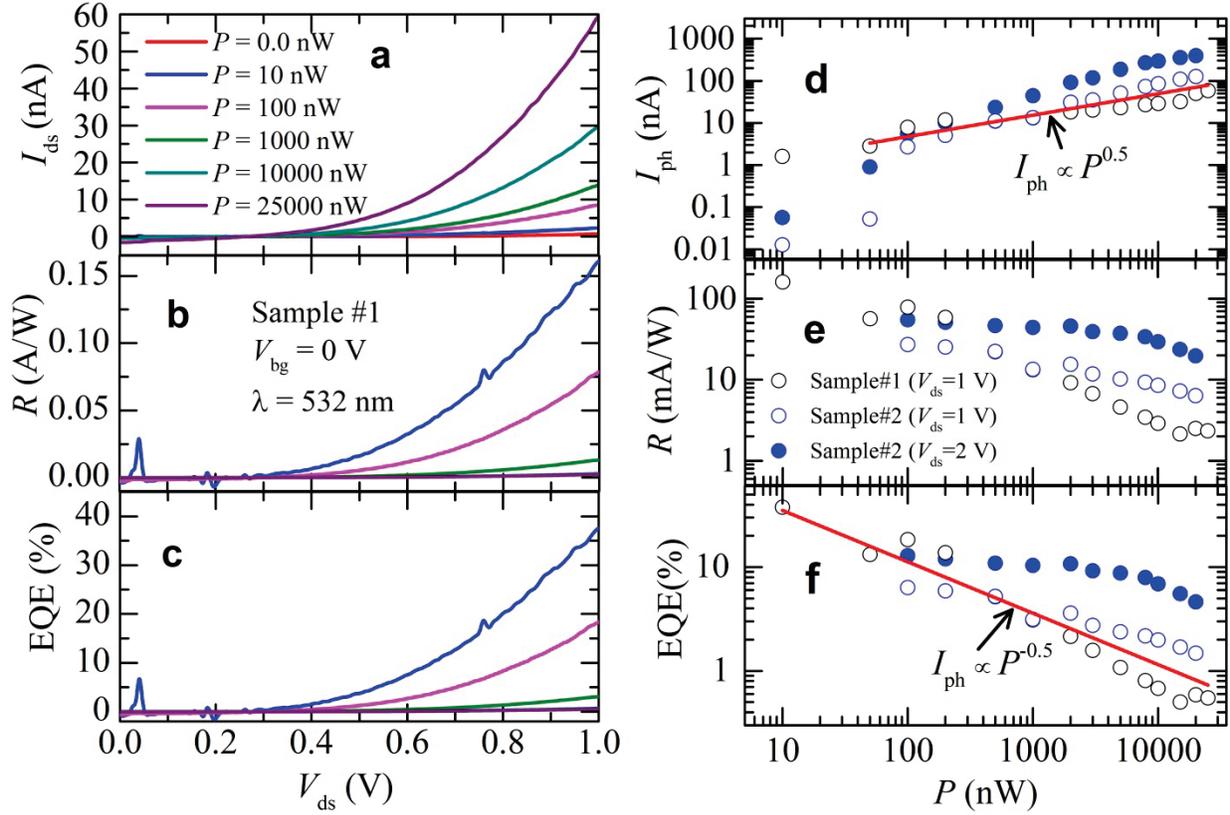

**Figure 4.** (a) Drain to source current $I_{ds}$ as a function of bias voltage $V_{ds}$ for a tri-layered WSe$_2$ field effect transistor under zero gate voltage ($V_{bg} = 0$ V) and for several values of the optical power (for a wavelength $\lambda = 532$ nm). (b) Photoresponsivity $R = I_{ph}/P$, where $I_{ph} = (I_{ds}(P) - I_{ds}(P = 0))$, as a function of $V_{ds}$. (c) Resulting external quantum efficiency EQE as a function of $V_{ds}$. (d) $I_{ph}$ as a function of the applied optical power $P$ in a log-log scale and for two samples, where open circles depict $I_{ph}(V_{ds} = 1$ V), and closed ones $I_{ph}$ for $V_{ds} = 2$ V (sample #2). Red line is a power law fit indicating an exponent $\alpha = 0.5$, or that $I_{ph}$ tends to saturate at high $P$ levels, likely due to an increased role for exciton recombination processes. (e) Photo responsivity as a function of laser power, $R(P)$, from the data in (d), indicating that $R$ reaches values as high as, or exceeds 100 mA/W at low $P$ levels, but decreases by several orders of magnitude as $P$ increases. (f) EQE as a function of $P$, in a log-log scale from the data in (e). (f) Red line is a linear fit yielding an exponent $\alpha = -0.5$.

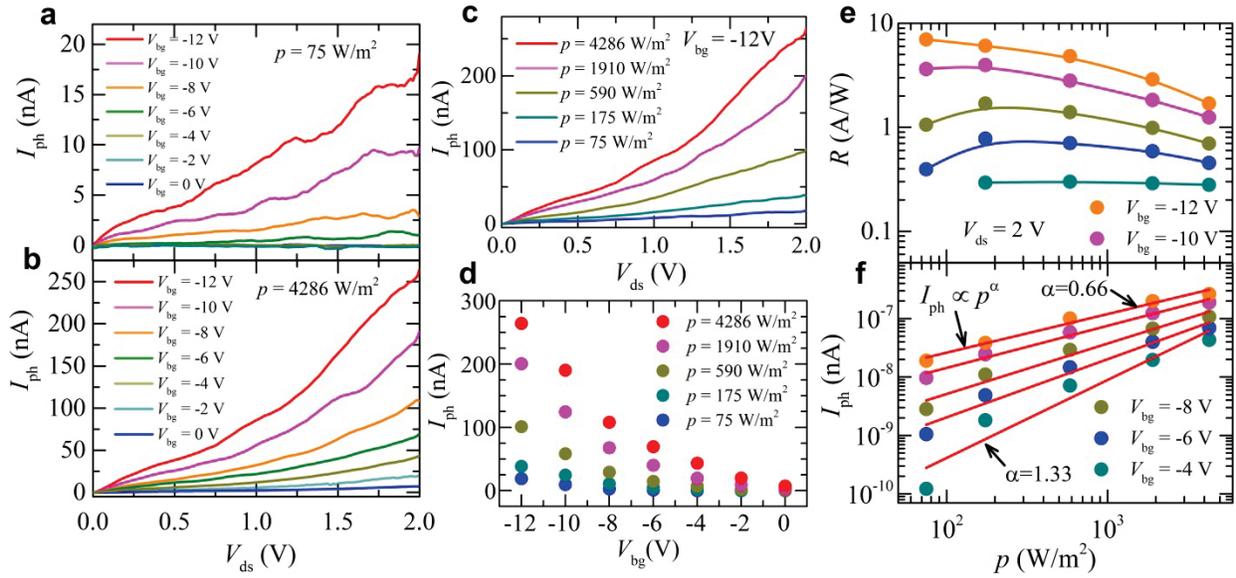

**Figure 5. (a)** Photo current $I_{ph}$ as a function of bias voltage $V_{ds}$ for a tri-layered WSe$_2$ field effect transistor, for several values of the gate voltage $V_{bg}$ under a fixed optical power density $p$ = 75 W/m$^2$ (white light spectrum produced of a Xe lamp). These measurements were performed on the same sample whose photo response under λ = 532 nm illumination is displayed in Fig. 4. **(b)** Same as in **(a)** but for $p$ = 4286 W/m$^2$. **(c)** $I_{ph}$ as a function of $V_{ds}$ under a constant gate voltage $V_{bg}$ = -12 V and for several $p$ values. **(d)** $I_{ph}$ as a function of the gate voltage $V_{bg}$ for several $p$ values. **(e)** Photoresponsivity $R$ as a function of the white illumination power density $p$ under a bias voltage $V_{ds}$ = 2 V and *for* several values of the gate voltage $V_{bg}$. Notice the remarkably high $R$ values at low $p$. **(f)** $I_{ph}$ as a function of $p$ in a log-log scale, under $V_{ds}$ = 2 V and for several values of the gate voltage $V_{bg}$. Red lines are linear fits, i.e. $I_{ph} \propto p^{\alpha}$, yielding an exponent α ranging from ~1.33 at low gate voltages to 0.66 at higher $V_{bg}$.

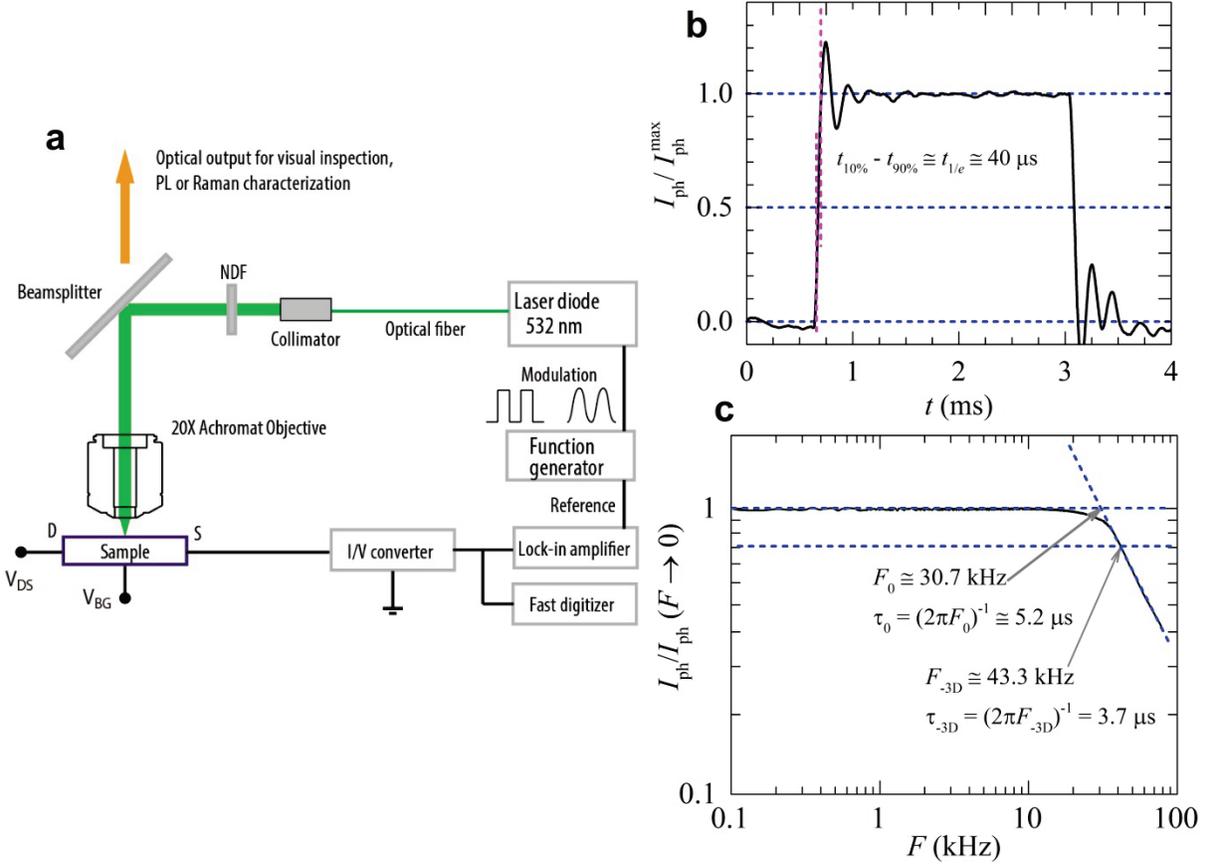

**Figure 6. (a)** Schematics of the experimental set-up for measuring time dependent photoresponse. **(b)** Photo-generated signal (normalized by its maximum value) as a function of time in the dark and under 532 nm laser illumination under a modulation of 400 Hz, a laser power of 7.9 µW measured at $V_{ds}$ = 1.5 V. The superimposed damped oscillatory component results from capacitive elements in the circuitry such as the gate capacitance. **(c)** Normalized photocurrent as a function of the laser modulation frequency.

**Figure: TOC**

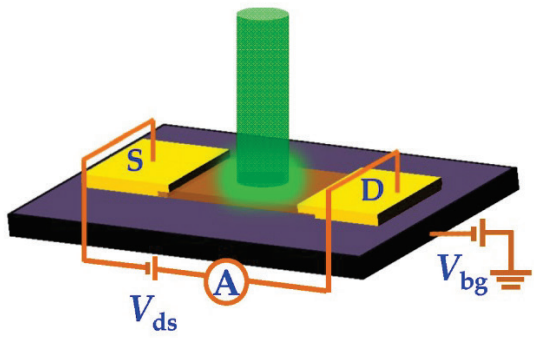
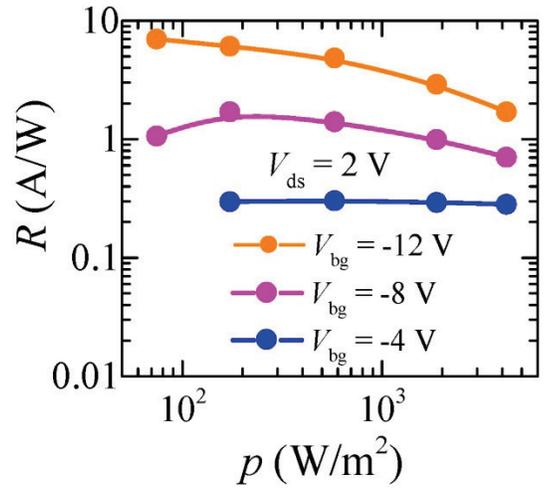
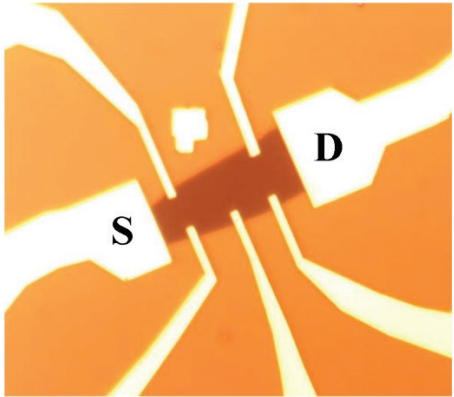
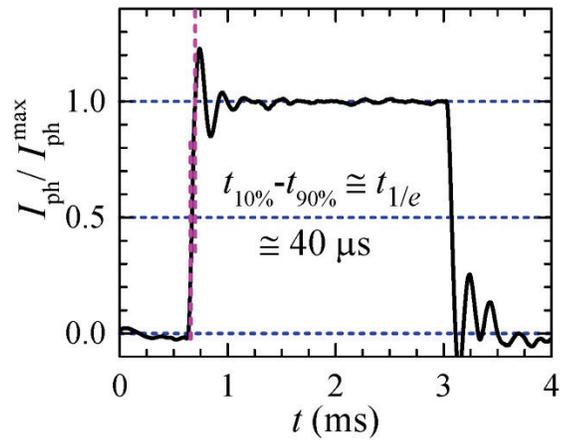

**Supporting Information** for manuscript titled: **"High Photoresponsivity and Short Photo Response Times in Few-Layered WSe$_2$ Transistors"** by Nihar R. Pradhan[1],* Jonathan Ludwig[1,2], Zhengguang Lu[1,2], Daniel Rhodes[1,2], Michael M. Bishop[1], Komalavalli Thirunavukkuarasu[1], Stephen A. McGill[1], Dmitry Smirnov[1,2] and Luis Balicas[1,2],*

[1]National High Magnetic Field Laboratory, Florida State University, Tallahassee-FL 32310, USA

[2]Department of Physics, Florida State University, Tallahassee, Florida 32306, USA

1. **Temperature dependent electrical transport characterization of tri-layered WSe$_2$ field-effect transistors.**

In Figure S1 below we present a detailed electrical transport characterization of mechanically exfoliated tri-layered WSe$_2$ field-effect transistors similar to the one used to evaluate the photoconducting response and as a function of the temperature $T$. For three atomic layers of WSe$_2$ mechanically exfoliated onto SiO$_2$ we observe ON to OFF ratios approaching $10^6$, at all $T$s and maximum field-effect mobilities surpassing 500 cm$^2$/Vs (see Figure S1**d**) at low temperatures. As seen through Figures S1**a**, **b**, and **c** the threshold gate voltage $V^t_{bg}$ for carrier conduction increases as the temperature is lowered. A priori two scenarios could explain the observed behavior, namely i) a sizeable Schottky barrier which would prevent carrier conduction *via* thermionic processes through the electrical contacts and ii) carrier localization due low dimensionality and disorder. As clearly indicated by Figure 3**a** in the main text, the current as a function of the voltage characteristics of our FETs is linear, thus indicating that thermionic or gate dependent thermionic processes, dominate the transport of carriers across the contacts. We evaluated the size of the Schottky barrier at the level of the contacts following thermionic field-emission theory, finding a nearly negligible barrier. Nevertheless, we argue that SiO$_2$ substrates

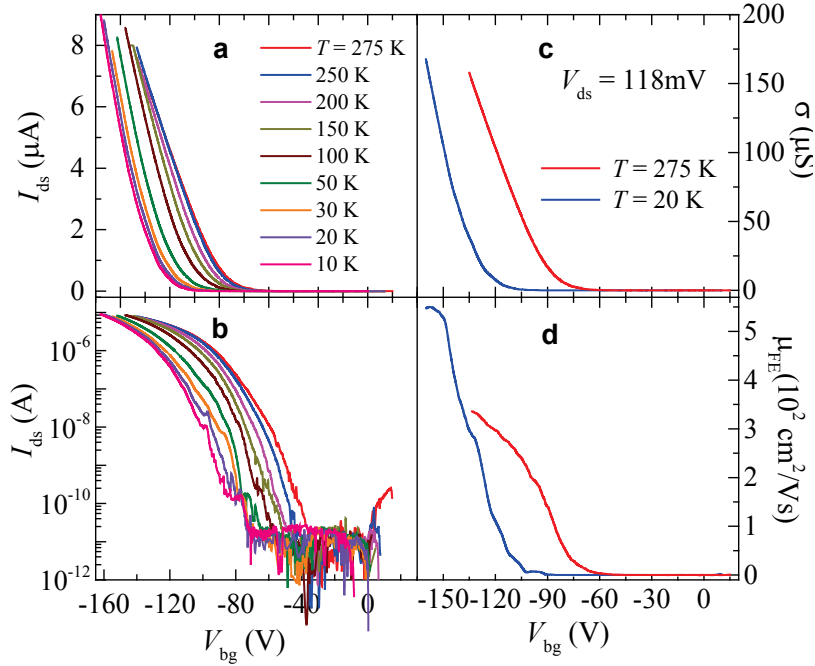

**Figure S1.** (a) $I_{ds}$ as a function of $V_{bg}$ under an excitation voltage $V_{ds}$ = 118 mV and for several temperatures. As seen the threshold gate voltage for carrier conduction continuously increases upon decreasing $T$. (b) same as in **a** but in a logarithmic scale. (c) Conductivity $\sigma = I_{ds}\ l/V_{ds}w$ where l and w are the length and the width of the channel, respectively for two temperatures ($T$ = 275 K and 20 K, respectively in red and in blue) and for an excitation voltage $V_{ds}$ = 118 mV. (d) Field-effect mobility $\mu_{FE} = (1/c_g)\ d\sigma/dV_{bg}$ where $c_g = \varepsilon_r\varepsilon_0/d$ = 12.789 x 10$^{-9}$ F/cm$^2$ (for a $d$ = 270 nm thick SiO$_2$ layer) as a function of gate voltage, as extracted from (**c**). At low temperatures we observed peak mobilities surpassing 550 cm$^2$/Vs.

(and possibly also lithographic residues from the sample fabrication process and adsorbates) are particularly detrimental to the performance of WSe$_2$: dangling bonds and charge traps lead to carrier localization and to a conductivity dominated by the two-dimensional variable range hopping transport mechanism as illustrated below by Figure S2.

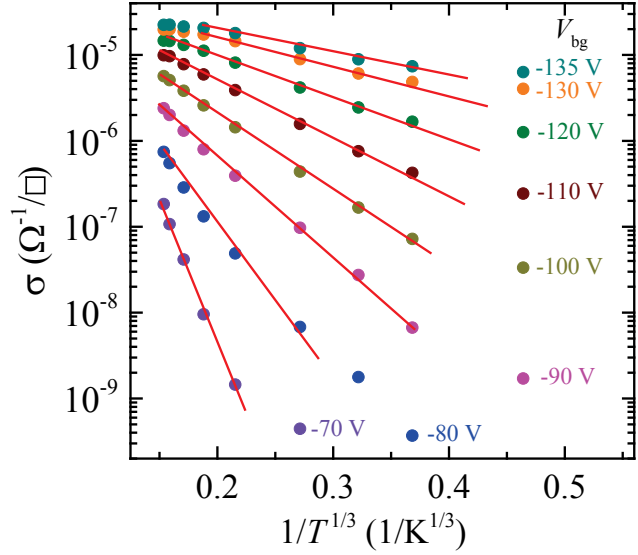

**Figure S2**. Two-terminal conductivity σ in a logarithmic scale and as a function of the inverse temperature to the 1/3 power or $T^{1/3}$. As seen, the expected two-dimensional variable range hopping dependence describes the conductivity data over an extended range of temperatures.

Therefore, one can safely conclude that the temperature dependence of the conductivity, which is dominated by an increase in the threshold gate $V_{bg}^t$ for carrier conduction is not entirely dominated by the role of the Schottky barriers at the level of the contacts. In order to understand the dependence of $V_{bg}^t$ on $T$, we will assume for a moment that $V_{bg}^t$ is dominated by disorder at the interface between the WSe$_2$ crystal and the SiO$_2$ layer which leading to charge localization. To illustrate this point, in Figure S2 we plot σ(T) as function of $T^{-1/3}$ since from past experience on Si/SiO$_2$ MOSFETs, it is well known that spurious charges[1] intrinsic to the SiO$_2$ layer[2,3] in addition to the roughness at the interface between the Si and the glassy SiO$_2$[S3] produces charge localization leading to variable-range hopping conductivity:

$$\sigma(T) = \sigma_0 \exp(-T_0/T)^{1/(1+d)} \qquad (S2)$$

where $d$ is the dimensionality of the system, or $d = 2$ in our case.[1,4] As seen in Figure S2, one observes a crossover from metallic-like (σ increases with decreasing $T$) to a clear two-dimensional variable-range hopping (2DVRH) conductivity below a gate voltage dependent temperature; red lines are linear fits. At lower gate voltages, the 2DVRH regime is observed over the entire range of temperatures. Therefore, and despite the linear (in current) transport regime and the relatively large mobilities displayed through Figures S2 and Figure 3 in the main text, this plot indicates very clearly, that below $V^t_{bg}$ the carriers in the channel are localized due to disorder. Notice that similar conclusions were also reported from measurements on $MoS_2$.[5] It is important however, to emphasize that we do not yet fully understand the type and role of disorder in these systems, which prevents us from developing any theoretical modeling.

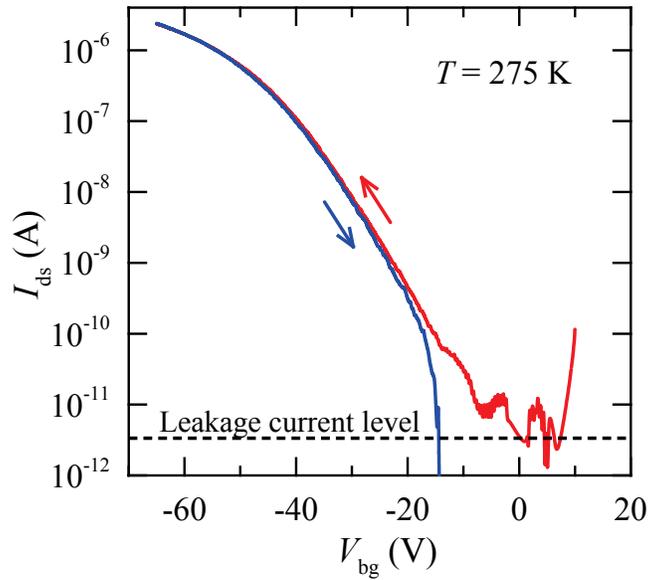

**Figure S3.** Drain to source current $I_{ds}$ as a function of the back-gate voltage $V_{bg}$ for a tri-layered $WSe_2$ based FET under a bias voltage $V_{ds}$ = 100 mV. Blue line depicts increasing gate voltage sweep, from negative towards the positive values, while the red one depicts a decreasing gate voltage sweep. Notice the very small hysteresis close to our leakage current level.

Now, we are in position of qualitatively explaining the $T$-dependence of $V^t_{bg}$: thermal activated processes promote carriers across a mobility edge which defines the boundary between extended electronic states and a tail in the density of states composed of localized electronic states. At higher temperatures, more carriers are thermally excited across the mobility edge, or equivalently, can be excited across the potential well(s) produced by disorder or charge traps, therefore one needs lower gate voltage(s) to un-trap the carriers. Once these carriers have moved across the mobility edge, they become mobile and, as our results show, respond linearly as a function of the excitation voltage $V_{ds}$. Finally, as $V^t_{bg}$ increases with decreasing $T$ the number of carriers is expected to decrease continuously since they become progressively localized due to the suppression of thermally activated processes which can no longer contribute to carrier detrapping.

## 2. Photocurrent and photoresponsivity for a 12 atomic layers thick WSe$_2$ field-effect transistor under white light illumination.

By increasing the number of atomic layers one expects to increase the probability of generating electron-hole pairs by incident photons, therefore one might expect a higher photoresponsivities. To test this hypothesis, we present a detailed characterization on a 12 atomic layer sample under the white spectrum produced by a Xe lamp.

As seen, at very low power densities one extracts similar photoresponsivities as the ones extracted for the tri-layered sample in the main text, i.e. approaching or exceeding 5 A/W at p < $10^2$ W/m$^2$, but with the application of a one order of magnitude smaller bias voltage, $V_{ds}$ = 0.1 V in contrast to 2 V for the tri-layered device in the main text.

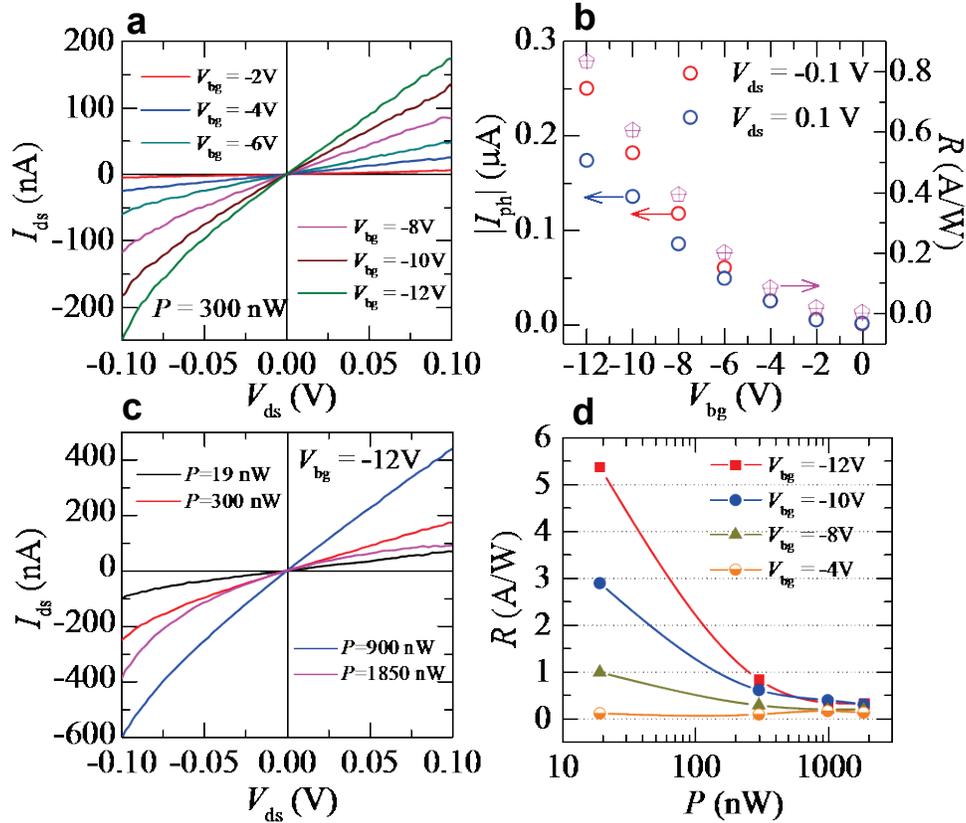

**Figure S4. (a)** Drain to source current $I_{ds}$ as a function of the bias voltage $V_{ds}$ for a 8 nm thick (~ 12 atomic layers) WSe$_2$ based FET under a white light illumination power $P = 300$ nW, and for several values of the gate voltage $V_{bg}$. **(b)** Absolute value of the generated photocurrent or $I_{ph}(P) = I_{ds}(P) - I_{ds}(P = 0)$, as a function of the gate voltage for two values of the excitation voltage $V_{ds} = 0.1$ (blue markers) and -0.1 V (red markers), respectively. The same figure also shows the corresponding photoresponsivity $R = I_{ph}/P$ (magenta markers) as functions of the back-gate voltage $V_{bg}$. **(c)** $I_{ds}$ as a function of $V_{ds}$ for several power levels under a gate voltage $V_{bg} = -12$ V. **(d)** $R$ as a function of the applied illumination power $P$ and for several values of Vbg. Notice the remarkably high values of $R$ approaching 5.5 A/W a low illumination power.

**Asymmetry in the photoresponse respect to the bias voltage and the role of the electrical contacts**

When performing the measurements displayed in figures 4 and 5 in the main text, we observed that the photocurrent response $I_{ph}(P)$ is asymmetric with respect to the sign of the applied excitation voltage $V_{ds}$, showing, for example, considerably larger values for $I_{ph}(V_{ds} > 0$ V) when compared to $I_{ph}(V_{ds} < 0$ V). In Figure S5 below we address this issue by demonstrating that the origin of this asymmetry is associated with the exact position, or more specifically with the relative proximity of the laser spot to one of the current contacts.

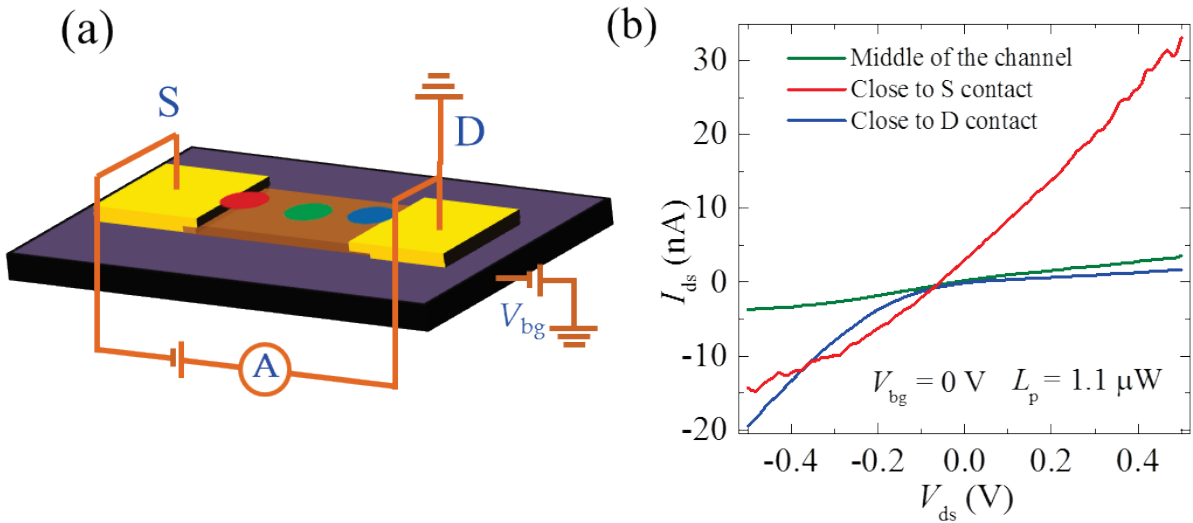

**Figure S5**. (a) Sketch of one of our WSe$_2$ phototransistors (in this case composed of a single atomic layer) when placed under illumination for photocurrent measurements. Red, green and blue spots depict the position of the laser spot relative to the current contacts, being located close to the drain contact, in the middle of the channel, and close to the source contact, respectively. (b) $I_{ds}$ as a function of $V_{ds}$ under illumination and when the laser spot is located in closer to source contact (red line), middle of the channel (green line), or closer to drain contact (blue line), As seen, for positive values of $V_{ds}$ one sees a clear increase in $I_{ds}$ (red trace) when the laser spot is placed on the drain contact. A similar situation is encountered when negative values of $V_{ds}$ are applied: one sees an increase in $I_{ds}$ when the laser spot is placed in proximity to the source contact. Therefore, under illumination the asymmetry in the $I_{ds}$ as a function of $V_{ds}$ traces can be attributed to some level of radiation acting on an area close to one of the contacts thus promoting the flow of carriers through it.[6]

The green traces in Fig. S5 depicting the drain to source current $I_{ds}$ as a function of the excitation voltage $V_{ds}$ were obtained with the laser spot carefully placed in the middle of the channel and act as a reference for the other traces obtained when the laser spot is displaced towards one of the current contacts. When the laser spot is closer to the drain (D) contact, one observes a pronounced photocurrent only for positive excitation voltages. The situation is reversed when the laser spot is displaced towards vicinity of the source contact, i.e. one obtains a sizeable current only for negative excitation voltages. As mentioned above explained in Ref.[6] this asymmetry is due to the role of the contacts that locally "bends" the conduction and valence bands, particularly under an excitation voltage. Hence, subsequent illumination in the vicinity of the contact area, generates electron-hole pairs which are accelerated to (holes) or from (electrons) the contacts due to the relatively large local electric fields. Inverting the excitation voltage bends the bands in the opposite way thus making it more difficult for holes and electrons to tunnel across the electrical contacts, and hence leading to comparatively much lower currents, as is effectively seen in our experiments.The important aspect here is the confirmation that the electrical contacts play a determinant role for the photo-response of field-effect transistors based on transition metal dichalcogenides. Therefore, this is an aspect that deserves a particular attention; by understanding the role of the electrical contacts one should be able to improve the considerably photoresponsivity of these systems.

3.Photocurrent, photoresponse and external quantum efficiency for a shorter excitation wavelength.

Figure S6 below shows the photocurrent results for a distinct and shorter wavelength, namely 408 nm, as well as the photoresponsivities and concomitant EQEs, respectively. Both R and EQE are lower than the values obtained for $\lambda = 532$ nm.

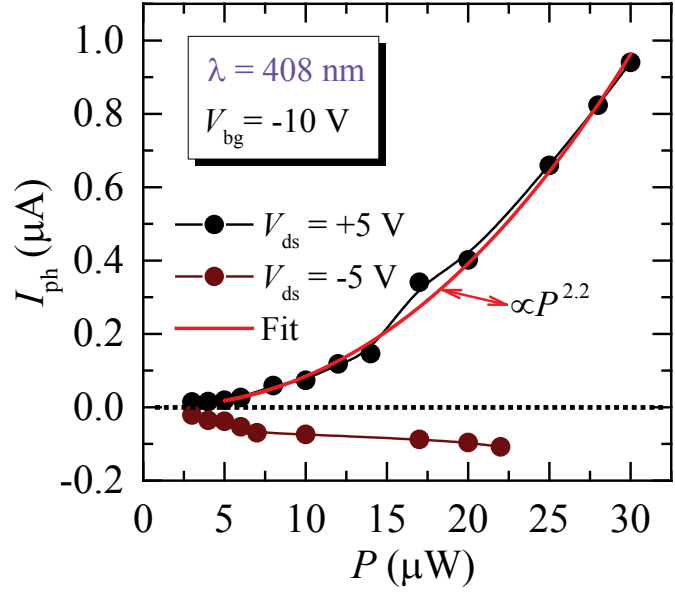

**Figure S6.** Photo-current, i.e. $I_{ph}(P) = I_{ds}(P) - I_{ds}(P=0)$ as a function of the incident laser power $P$, for a gate voltage of -10 V and for two values of the excitation voltage, or respectively $V_{ds} = \pm 0.5$ V under an illumination wavelength $\lambda = 408$ nm. Red line is a power law fit $I_{ph} = P^{\alpha}$ with $\alpha \approx 2.2$ thus indicating that the photothermoelectric effect, which would yield $\alpha \approx 1$, is not responsible for the observed photoresponse. We extract an average photo responsivity of 20 mA/W and average EQE of ~ 6 %.